\documentclass[10pt,letterpaper,twocolumn]{article}

\usepackage{ol}
\usepackage{hyperref}
\usepackage{amsmath,amssymb}

\newcommand{\figwidth}{0.442\textwidth}

\begin{document}

\twocolumn[

\newcommand{\ethaffil}{Laboratory of Physical Chemistry, Swiss Federal Institute of Technology (ETH), 8093 Zurich, Switzerland}
\newcommand{\mariaaffil}{IESL, Foundation for
Research and Technology Hellas (FORTH), P.O. Box 1527, 71110
Heraklion, Crete, Greece}

\title{Spontaneous emission in the near-field of 2D photonic crystals}

\author{A. Femius Koenderink}
\affiliation{\ethaffil}
\author{Maria Kafesaki}
\affiliation{\mariaaffil}
\author{Costas M. Soukoulis}
\affiliation{Ames Laboratory, Iowa State University, Ames, Iowa
50011, and IESL, FORTH, P.O. Box 1527, 71110 Heraklion, Greece,
and Dept. of Materials Science and Technology, University of
Crete, Greece }
\author{Vahid Sandoghdar}
\affiliation{\ethaffil}

\date{Submitted July 13, 2005 }

\begin{abstract}
We show theoretically that photonic crystal membranes cause large
variations in the spontaneous emission rate of dipole emitters,
not only inside but also in the near-field above the membranes.
Our three-dimensional finite difference time-domain calculations
reveal an inhibition of more than five times and an enhancement of
more than ten times for the spontaneous emission rate of emitters
with select dipole orientations and frequencies. Furthermore we
demonstrate theoretically, the potential of a nanoscopic emitter
attached to the end of a glass fiber tip as a  local probe for
mapping the large spatial variations of the photonic crystal local
radiative density of states. This arrangement is promising for
on-command modification of the coupling between an emitter and the
photonic crystal in quantum optical experiments.
\end{abstract}


]

It is well known that the rate of spontaneous emission can be
controlled by the geometry of the medium surrounding  the
fluorescent species. In particular, many recent research efforts
have been devoted to studying spontaneous emission in photonic
crystals\cite{soukou01,lodahl04,ogawa04}. The quantitative
interpretation of these experiments, however, remains frustrated
by lack of detailed information about many parameters that
strongly affect the emission dynamics. These include the exact
position of the emitters on the subwavelength scale and the
orientation of the emission dipole moments, as well as systematic
effects such as surface-induced quenching\cite{quench} or other
chemical or electronic surface phenomena. An ideal arrangement
would require accurate placement of a single emitter at an
arbitrary location in a photonic crystal (PC). Very recently
Badolato \emph{et al}. have achieved this by precise fabrication
of a PC structure around a given semiconductor
emitter\cite{badolato}. In this Letter we discuss the
\emph{in-situ} control of the position and thereby modification of
the spontaneous emission rate of a single emitter close to or in a
two-dimensional PC slab.

Two-dimensional (2D) photonic crystals fabricated in thin
semiconductor membranes promise to achieve many of the
long-standing goals of photonic band gap materials. Indeed,
recently it has been demonstrated that it is possible to achieve
very high-Q and low mode volume cavities in these
structures.\cite{yoshie,akahane03} Due to their planar nature, PC
membranes can be easily accessed by subwavelength probes such as
optical fiber\cite{kramper} or atomic force microscope
tips\cite{nanoswitch}. Motivated by this opportunity, we
investigate the prospects of coupling between a PC and nanoscopic
optical emitters located at the end of sharp
probes\cite{michaelis,tipemit,zwiller}. Although 2D crystals do
not yield a zero density of states, we show that both inside and
in the near-field above a PC membrane the emission rate of
properly oriented dipoles can be strongly modified. We show that
the nanometer accuracy in scanning probe positioning allows direct
mapping of the dependence of the emission rate on the spatial
coordinates of the subwavelength emitter.

We have used the three-dimensional Finite-Difference Time-Domain
(FDTD) method\cite{lee,taflove,hermann} to calculate the local
radiative density of states (LRDOS), accounting for the position
dependence of the photon states available for fluorescent decay of
a quantum emitter\cite{sprik96}. This calculation relies on the
fact that the LRDOS appearing in the formulation of Fermi's Golden
Rule for the spontaneous emission rate, also describes the total
emitted power of a classical point-dipole antenna run at a fixed
current\cite{lee}. We consider semiconductor membranes with
dielectric constant $\epsilon=11.76$ and thickness $d=250$~nm,
surrounded by up to $1~\mu$m of air above and below. We take the
membrane to contain a hexagonal array of holes with radius
$r=0.3a$ at a lattice spacing of $a=420$~nm. Such a structure
possesses a band gap in the range $a/\lambda=0.25$ to 0.33 for the
transverse electric (TE) mode where the electric field is parallel
to the plane of the membrane. The ratio $a/\lambda$ is used as
normalized frequency units throughout our work. We used
discretization with 14 or 20 points per lattice constant and
employed volume-averaging of the dielectric constant to reduce
staircasing errors.\cite{nanoswitch} We considered finite
hexagonally shaped PC structures  up to $25$ holes across,
terminated by the unperforated slab extending into Liao's
absorbing boundary conditions. By broadband temporal excitation of
the dipole, we simulated the emission power spectrum over a wide
frequency range. After dividing the resulting spectrum by that of
an identically excited dipole in vacuum, we obtained the LRDOS
normalized to the vacuum LRDOS~\cite{hermann}.

\begin{figure}[t]
\centerline{\includegraphics[width=\figwidth]{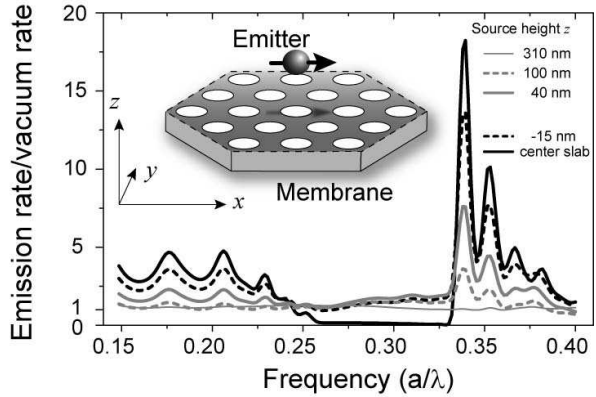}}
 \caption{\label{fig:fig1}%
Emission rate normalized to the vacuum rate versus frequency for
an $x$-oriented dipole in the central hole of a PC membrane
(details in text). Black lines correspond to dipoles in the slab
($z<0$) and gray lines to dipoles above the slab ($z>0)$, as
listed in the legend.}
\end{figure}

For an emitter half way deep in a PC membrane, the solid black
spectrum in Fig.~\ref{fig:fig1} shows a strong inhibition of
fluorescence by over a factor of 7 in the band gap as compared to
its vacuum rate. In this case the dipole was laterally centered in
the structure, and its orientation was chosen in the $x$-direction
(cf. Fig.~\ref{fig:fig1}). The slab was taken to be $13a$ across,
and we verified that no significant further reduction of the
emission rate was obtained if we increased the size of the
structure. However, the magnitude of the enhancement at the blue
edge of the gap, as well as the Fabry-P\'erot oscillations at
frequencies below the gap depend on the finite size of the PC
structure. For all tested structures wider than 7 holes across, we
find emission enhancements larger than a factor of 15,
representing a jump of two orders of magnitude as compared to the
LRDOS for frequencies in the gap.

\begin{figure}[t]
\centerline{\includegraphics[width=\figwidth]{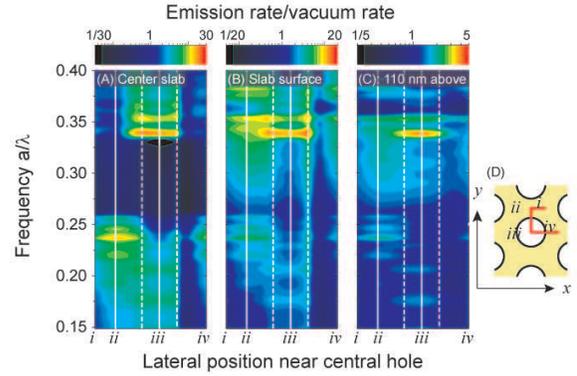}}
\caption{\label{fig:fig2}(color) Emission rate normalized to the
vacuum rate for an $x$-oriented dipole (a) in the mid-depth, (b)
on the surface and (c) 110 nm above the PC membrane as a function
of frequency and position along the trajectory indicated by the
red line in (d). The trajectory traces the edges of the
irreducible part of the unit cell. The dashed lines mark the
borders between air hole and dielectric. The logarithmic color
scales are shown on top.}
\end{figure}

Given the inherent strong modulation of the dielectric constant in
a PC structure, it is particularly interesting to examine the
lateral dependence of the spontaneous emission rate.
Figure~\ref{fig:fig2}(a) shows a contour plot of the LRDOS
modification for an $x$-oriented dipole midway in the slab depth
versus emission frequency and for lateral locations along a
trajectory that traces the irreducible part of the unit cell
(Fig.~\ref{fig:fig2}(d)). The emission is inhibited in the band
gap at all positions whereas outside the gap we observe
Fabry-P\'erot modulations together with enhancement at the low and
high frequency edges. The enhancement of the emission   occurs
especially on the high frequency edge of the gap (the `air band')
for dipoles   in   air holes, and predominantly at the low
frequency edge (the `dielectric band') for dipoles in the
dielectric.

\begin{figure}[t]
\centerline{\includegraphics[width=\figwidth]{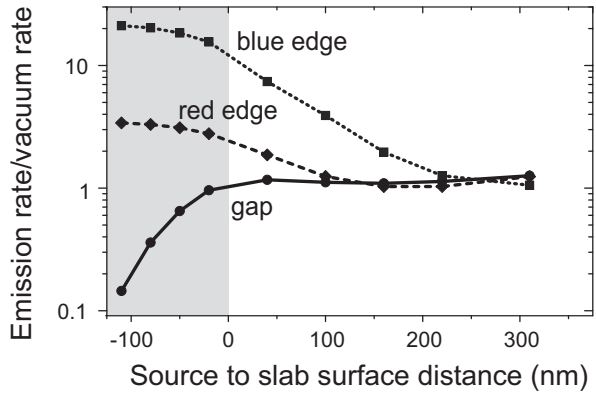}}%
\caption{\label{fig:fig3}%
Emission rate modification as a function of the height of a dipole
above the PC membrane. Circles, diamonds and squares show data for
$a/\lambda=0.23,0.28,0.34$, corresponding to frequencies below, in
and above the gap, respectively. The shaded region shows the range
of positions $-125<z<0$~nm in the membrane.}
\end{figure}

Next, we ask whether it is possible to capture these effects by
scanning an emitter just above the PC slab. Different spectra in
Fig.~\ref{fig:fig1} show the LRDOS modification of a dipole
laterally centered in the structure but at various heights $z$
above the membrane. In addition Fig.~\ref{fig:fig2}(b) and (c)
display the modification of the LRDOS for the dipole right at the
crystal-air interface and at $110$~nm above this plane. These data
reveal that as $z$ increases, the inhibition and enhancement
reduce in size. To examine this distance dependence more closely,
in Fig.~\ref{fig:fig2} we plot the normalized emission rate as a
function of the distance between the dipole and the membrane
surface for three key frequencies $a/\lambda=0.23, 0.28$ and
$0.34$ just below, in, and just above the band gap, respectively.
Evidently, the inhibition diminishes for emitters located above
the slab. In contrast, enhancements persist at the blue edge of
the gap even if the dipole is lifted into air above the membrane.
Figures~\ref{fig:fig2} and~\ref{fig:fig3} let us conclude that it
is possible to enhance the spontaneous emission rate by a factor
of $5$ to $10$ if the emitter position is controlled to within
$50$~nm above the PC membrane. Note that the emission of a dipole
near a simple homogeneous dielectric slab is also enhanced, due to
coupling to the guided modes. However, at the gap edges  the PC
causes a further strong enhancement of the
LRDOS.\cite{nanoldoslong}

The required resolution and control for mapping the modification
of the emission rate can be achieved by scanning a subwavelength
emitter attached to the end of a sharp
tip\cite{michaelis,tipemit,zwiller}. A crucial question arises as
to the effect of the tip on the LRDOS. To estimate this effect, we
have calculated the LRDOS for sources embedded inside cylindrical
tips of diameter 125~nm pushed 130~nm into the central hole  of
the PC structure. We find that the influence of the PC structure
on the spectrum of LRDOS enhancement and inhibition is unchanged
for the system of emitters embedded in glass ($\epsilon\lesssim
2.25$). In contrast, tips of very high-index material such as
silicon fundamentally change the LRDOS spectrum, leading to the
creation of a low-Q localized defect mode from the band edge due
to the addition of dielectric material.\cite{nanoswitch} The
presence of a silicon tip causes an overall reduction of the gap
depth and an increase and red-shift of the rate enhancement at the
blue edge of the gap.  Suitable probes of the LRDOS in photonic
crystal membranes are therefore emitters inside low index tips.

Although optical detection of single emitters has become possible
for some systems,  many applications such as the realization of a
nanolaser would benefit from coupling  an ensemble of emitters to
a PC. Furthermore, nanoscopic ensembles are more readily available
than  single emitter systems. Thus, we have also considered the
modification of the spontaneous emission rate for a subwavelength
ensemble of randomly oriented dipoles. Note that because the LRDOS
is essentially unchanged for the TM polarization, dipolar
components normal to the slab reduce the visibility of lifetime
effects. We have considered over 25 in-equivalent dipole
orientations (corresponding to over 300 orientations in $2\pi$
solid angle) in the central air hole and have calculated the
corresponding LRDOS and luminescence extraction
efficiency.\cite{lee,nanoldoslong,fanextract} We find that in
general, the time-resolved flux of fluorescence photons extracted
from the slab follows a significantly non-single exponential decay
behavior. Nonetheless, the mean decay constant reveals inhibition
by a factor three, and enhancement by a factor five compared to
vacuum.  The observation of inhibition is facilitated by the
increase of the emission extraction efficiency for in-plane
dipoles from $\sim 20\%$ for frequencies below the gap to $>80\%$
in the gap.\cite{lee,fanextract}

In conclusion, we have shown that   strong inhibition and
enhancement of emission can be achieved for emitters well inside
photonic crystal membranes while a significant level of
enhancement persists   even in the near field above the
structures. Since these results also hold for emitters embedded in
nanoscopic dielectric probes,
 scanning probe technologies can be promising for
on-command  spontaneous emission control. An important advantage
of  emitters inside such probes is that they  are shielded from
unwanted interactions,  and  can be calibrated by simply
retracting the probe from the structure.

This work was funded by the Deutsche Forschungsgemeinschaft (DFG)
through focus program SP1113 and by ETH Z\"urich. Vahid Sandoghdar
can be reached by e-mail at vahid.sandoghdar@ethz.ch.

\end{document}